\documentclass[preprint,preprintnumbers,amsmath,a4paper,aps,12pt]{revtex4}

\usepackage{bm}
\usepackage{graphicx}
\newcommand{\BEA}{\begin{eqnarray}}
\newcommand{\BEAN}{\begin{eqnarray*}} 
\newcommand{\EEA}{\end{eqnarray}}
\newcommand{\EEAN}{\end{eqnarray*}}

\newcommand{\beq}{\begin{equation}}
\newcommand{\eeq}{\end{equation}}
\newcommand{\threejm}[6]{ \left(\begin{array}{ccc} #1 & #3 & #5\\
                                              #2 & #4 & #6
                                \end{array}
                          \right)}

\begin{document}

\preprint{APS}

\title{Ultracold $O_2$+$O_2$ collisions in a magnetic field: on the role of the
  potential energy surface}


\author{Jes{\'u}s P{\'e}rez-R\'{\i}os,   
Jos\'e Campos-Mart\'{\i}nez, Marta I. Hern\'andez\footnote{Electronic mail: 
marta@iff.csic.es}}
\affiliation{Instituto de F\'{\i}sica Fundamental, 
Consejo Superior de Investigaciones Cient\'{\i}ficas, Serrano 123, 
28006 Madrid, Spain}

\date{\today}

\begin{abstract}
The collision dynamics of $^{17}O_2(^3\Sigma_g^-)
+^{17}O_2(^3\Sigma_g^-)$ in the presence of a magnetic field is studied within
the close-coupling formalism in the range between 10 nK and 50 mK. 
A recent global {\em ab
  initio}  potential energy surface (PES) is employed and its effect on the
dynamics is analyzed and compared with previous calculations where an
experimentally derived PES was used [New J. Phys {\bf 11}, 055021 (2009)].  
In contrast to the results using the older PES, magnetic field
dependence of the low-field-seeking state in the ultracold regime is 
characterized by quite a large background scattering length, $a_{bg}$,
and, in addition, cross sections exhibit 
 broad and pronounced Feshbach resonances. The marked resonance
 structure is somewhat surprising considering the influence of 
inelastic scattering, but it can be explained by resorting to the analytical van
der Waals theory, where the short range amplitude of the entrance channel wave
function is enhanced by the large $a_{bg}$. This strong sensitivity to the
short range of the {\em ab initio} PES persists up to relatively high energies (10 mK). 
After this study and despite quantitative predictions are very difficult, it
can be concluded that the ratio between elastic and spin relaxation scattering
is generally small, except for magnetic fields which are either low 
 or close to an asymmetric Fano-type resonance. 
Some general trends found here, such as a large density of quasibound states
and a propensity towards large scattering lengths, could  be also characteristic of
other anisotropic molecule-molecule systems.
\end{abstract}

\maketitle

\section{Introduction}

Ultracold  molecules play a main role in modern
physics due to a large number of promising applications in quantum
information\cite{ref1-Krems}, precision spectroscopy\cite{ref3-Krems} and
ultracold chemistry\cite{ref4-Krems,Softley:09,Gianturco:09}.  
Optical lattices of ultracold molecules are
predicted to be ideally suited for quantum simulation of complex 
quantum systems\cite{ref2-Krems,ref8-Krems,ref5-NJP} and the engineering
of new schemes for quantum information storage and
processing\cite{ref6-Krems,ref7-NJP}. On the other hand, creation 
 of a Bose-Einstein condensate (BEC) of molecules may 
enable studies of Bose-enhanced chemistry\cite{ref10-Krems}.
In the context of these studies, molecules must be confined within a trap.
For paramagnetic molecules, a magnetic trap is used since molecules
in a low-field-seeking state\cite{Pethick} ({\em lfs}) are
trappable provided that their translational energy 
 is lower than the trap depth\cite{Freidrich-Doyle}. This situation
  could be achieved by  direct cooling methods
 such as Zeeman slowing \cite{ref37-Krems}, optical Stark deceleration
 \cite{ref56-Krems}, single-photon cooling\cite{Raizen-single-photon} 
or sympathetic cooling\cite{Salomon:01}.
It might also be possible to cool the molecules towards the ultracold regime  by
evaporative cooling\cite{Hess}. As it is well known, this was the successful method for
achievement of BEC of atoms\cite{Wieman-BEC,Ketterle-BEC}.

Molecular collisions are fundamental in this context, as evaporative cooling
relies on efficient elastic collisions and, even more crucially, on the ratio of the
probabilities for elastic scattering and spin relaxation ($\gamma$) which must be
very large in order to prevent  heating and trap loss. External
electromagnetic fields may serve to control the rate of inelastic collisions.  
Tuning close to a Feshbach resonance has proved to be an extremely fruitful
means of controlling atom-atom collisions\cite{Chin:10}. Interestingly, 
it has been recently shown\cite{Hutson:09} that inelastic collision rates in
atom-molecule collisions can
be tremendously reduced in the vicinity of a Feshbach resonance controlled by
an electric or magnetic field.

While a large amount of work has been carried out for atom-atom and
atom-molecule collisions, studies of molecule-molecule collisions in external
fields are still scarce. Most clues about these more complex systems have come
from atom-molecule studies. Krems and Dalgarno\cite{ref20-Krems} found that
the main mechanisms of spin relaxation in collisions of $^3\Sigma$ molecules
with He is given by couplings to rotationally excited states mediated by the
spin-spin interaction. Volpi and Bohn\cite{ref24-Krems} found that spin
depolarization is suppressed when the Zeeman splitting between incident and final
states does not exceed the height of the centrifugal barrier in the exit
channel. These ideas were confirmed for $^{17}O_2(^{3}\Sigma_{g}^{-})$ +
$^{17}O_2(^{3}\Sigma_{g}^{-})$  by  Tscherbul {\em et
  al}\cite{paper-Krems}, who carried out the first accurate computational
study involving two diatoms. In that work, the
experimentally derived potential energy surface (PES) of Aquilanti {\em et al}\cite{Perugia-PES} was
employed  (Perugia PES in what follows). This collisional system is
interesting since oxygen has been postulated as a
reliable candidate for trapping and cooling\cite{Friedrich,ref25-Krems} and
progress in cooling this species has been
achieved recently\cite{ref37-Krems,ref38-Krems}. 

The present work builds up along these lines by the investigation of the role
played by the PES in $O_2$+$O_2$ collisions in the
presence of a magnetic field. It is well known that
ultracold atom-atom collisions are very sensitive to the short range of the
potential\cite{Gribakin:93}. However, it has been recently
shown\cite{Hutson:07} that, in the presence of inelastic scattering
(i.e. atom-molecule collisions), peaks in cross sections around a Feshbach
resonance may become suppressed and hence dynamics becomes rather insensitive to the
details of the potential. This theory is tested here for a
rather anisotropic molecule-molecule system such as $O_2$ + $O_2$, using 
 a recent {\em ab initio} PES developed by Bartolomei {\em
  et al}\cite{abi-PES}. In this potential, electronic
correlation is included by means of a high level supermolecular method in the
short range whereas long-range interaction coefficients have been obtained from first
principles as well\cite{long-range}. It is worthwhile to mention that inelastic
rate coefficients obtained with this PES have proved to be highly consistent with
measurements 
of the evolution of rotational populations along supersonic
expansions in the temperature range 10 $\le T \le$ 34 K\cite{Montero:11}. 
 By comparing present scattering calculations with previous ones using 
 the Perugia PES\cite{paper-Krems} and with some additional test
 modifications of the {\em ab initio} PES, the effect of the potential on
 the cold and ultracold dynamics has been assessed. 

The paper is organized as follows. In Sec. II, a summary of the theory for the
scattering between two identical $^3\Sigma$ molecules is given. Details
specific to the $^{17}O_2-^{17}O_2$ system are provided in Sec. III and in
Sec.IV, results are reported and discussed. A concluding remark is given in
Sec. V.

\section{Theory}

We give a summary of the theory -recently developed by Tscherbul
{\em et al}\cite{paper-Krems}- for the scattering of two $^{3} \Sigma$ identical
rigid rotor molecules in the presence of a magnetic field. 
Diatom-diatom Jacobi coordinates are used in a space-fixed (SF) frame, including the
vector joining the centers of mass of the molecules $a$ and $b$, $\vec{R}$,
and the intramolecular unit vectors, $\hat{r}_{a}$ and
$\hat{r}_{b}$. Intramolecular distances are fixed to the molecular equilibrium
distance, $r_a=r_b=r_e$. The Hamiltonian for the interaction can be written as  

\begin{equation}
\label{ec1}
\hat{H}=-\frac{1}{2\mu R} \frac{\partial^{2}}{\partial R^{2}} R +
\frac{\hat{l}^{2}}{2\mu R^{2}}+V(\vec{R},\hat{r}_{a},\hat{r}_{b})+ \hat{H}_{a}
+ \hat{H}_{b}, 
\end{equation}

\vspace{.2cm}

\noindent
where atomic units are used ($\hbar=1$), $\hat{l}$ is the orbital angular
momentum, $\mu$ is the reduced mass and
$V$ is the interaction potential or PES. The
internal Hamiltonian of the $^{3} \Sigma$ molecule $\hat{H}_{\alpha} (\alpha=
a,b)$ is given, within the rigid rotor approximation, by\cite{Mizushima} 

\begin{eqnarray}
\label{ec2}
\hat{H}_{\alpha} & = & B_{e}\hat{n}_{\alpha}^{2}+2\mu_{B}\vec{B} \cdot
\hat{s}_{\alpha}+\gamma_{sr}\hat{n}_{\alpha} \cdot \hat{s}_{\alpha}+ \\ \nonumber
 & & +
\frac{2}{3}\lambda_{ss}\sqrt{\frac{24\pi}{5}}
\sum_{q}Y_{2q}^{*}(\hat{r_{\alpha}})[\hat{s}_{\alpha}\otimes \hat{s}_{\alpha}]^{(2)}_{q}, 
\end{eqnarray}

\vspace{.2cm}

\noindent
where $\hat{n}_{\alpha}$ is the angular momentum associated with
$\hat{r}_{\alpha}$,  $B_{e}$ is the rotational constant, $\mu_{B}$ is the Bohr
magneton, $\vec{B}$ is the external magnetic field  and $\hat{s}$ is the
electron spin. The last two terms in Eq.\ref{ec2} correspond to the
spin-rotation and spin-spin interactions, parameterized by 
$\gamma_{sr}$ and $\lambda_{ss}$, respectively.  
Weaker interactions such as hyperfine and magnetic dipole-dipole are neglected
(see Ref.\cite{ref25-Krems} for discussion).

The total wave function is expanded in a basis set of SF
uncoupled and symmetry-adapted functions 

\begin{equation}
\label{ec3}
\Psi^{M \eta \epsilon}=\frac{1}{R}\sum_{\tau_{a} \ge \tau_{b} l m_{l}}
u^{M\eta\epsilon}_{\tau_{a}\tau_{b}lm_{l}}(R) \phi^{M \eta \epsilon}_{\tau_{a}\tau_{b}lm_{l}}
(\hat{R},\hat{r}_{a} \hat{r}_{b}), 
\end{equation}

\vspace{.2cm}

\noindent
with

\begin{equation}
\label{ec4}
\phi^{M \eta \epsilon}_{\tau_{a}\tau_{b}lm_{l}} = 
\frac{1}{\left(2\left( 1+\delta_{\tau_{a},\tau_{b}}\right) \right)^{1/2}} 
 \left( |\tau_{a}\tau_{b}\rangle+\eta \epsilon |\tau_{b}\tau_{a}\rangle
 \right)  |l m_l\rangle, 
\end{equation}

\vspace{.2cm}

\noindent
$|l m_l\rangle$ being a spherical harmonics and where $| \tau_{\alpha}
\rangle$ represents an uncoupled function of the $\alpha$ monomer

\begin{equation}
\label{ec5}
| \tau_{\alpha} \rangle = |n_{\alpha} m_{n_{\alpha}} \rangle |s_{\alpha}
m_{s_{\alpha}}\rangle. 
\end{equation}

\vspace{.2cm}

\noindent
The basis of Eq.\ref{ec4} are a well-ordered set with $\tau_{a} \ge \tau_{b}$
that are normalized eigenfunctions of the
operator permuting the identical molecular skeletons
($\hat{P}$:  $\hat{r}_{a}\rightarrow \hat{r}_{b}$; $\hat{r}_{b}\rightarrow \hat{r}_{a}$;
$\vec{R}\rightarrow -\vec{R}$), with eigenvalue $\eta$.
 These basis functions are also eigenfunctions of
spatial inversion ($E^{*}$: $\hat{r}_{a}\rightarrow -\hat{r}_{a}$;
$\hat{r}_{b}\rightarrow -\hat{r}_{b}$;  $\vec{R}\rightarrow -\vec{R}$) with
eigenvalue $\epsilon= (-1)^{n_a+n_b+l}$. Since the molecules
under study are homonuclear, $n_a$ and $n_b$ have the same parity 
so $\epsilon=(-1)^{l}$. In addition to these symmetries, the Hamiltonian
commutes with the SF $z$-axis component of the total
angular momentum, so that for a given value of this projection, $M$, 
basis functions in Eq.\ref{ec3} must satisfy

\begin{equation}
\label{ec6}
m_{n_a}+m_{s_a}+ m_{n_a}+m_{s_a}+ m_l = M.
\end{equation}

\vspace{.2cm}

Substitution of Eq.\ref{ec3} into the Schr{\"o}dinger equation leads to the
set of close-coupled equations for the radial coefficients:

\begin{eqnarray} 
\label{ec7}
\left[ \frac{1}{2\mu} \frac{d^{2}}{d R^{2}}-\frac{l(l+1)}{2\mu R^{2}}+E\right]
u^{M \eta \epsilon}_{\tau_{a}\tau_{b}lm_{l}}(R) 
& = & 
\end{eqnarray}
\begin{eqnarray} 
\hspace{-.4cm} & = & \hspace{-.6cm} \sum_{\tau'_{a} \ge \tau'_{b} l' m'_{l}}  \hspace{-.4cm}
 \langle \phi^{M \eta \epsilon}_{\tau_{a}\tau_{b}l m_{l}} | 
(V+ \hat{H}_{a} + \hat{H}_{b}) | \phi^{M \eta \epsilon}_{\tau'_{a}\tau'_{b}l'm'_{l}} \rangle
u^{M\eta\epsilon}_{\tau'_{a}\tau'_{b}l'm'_{l}}(R), \nonumber
\end{eqnarray}

\vspace{.2cm}

\noindent
where $E$ is the total energy. 
It must be pointed out that the asymptotic Hamiltonian $\hat{H}_{a} +
\hat{H}_{b}$ is not diagonal in the basis $\phi^{M
  \eta\epsilon}_{\tau'_{a}\tau'_{b}l'm'_{l}}$ due to the spin-rotation and
spin-spin terms, and matrix elements of these terms are given
in Eqs.14 and 16 of Ref.\cite{ref20-Krems}, respectively. 
On the other hand, potential matrix elements are
given as a sum of a direct and an exchange coupling terms\cite{paper-Krems}: 

\begin{equation}
\langle \phi^{M \eta \epsilon}_{\tau_{a}\tau_{b}lm_{l}} | V |
        \phi^{M \eta \epsilon}_{\tau'_{a}\tau'_{b}l'm'_{l}} \rangle = 
\frac{1} 
{ [(1+\delta_{\tau_{a},\tau_{b}})(1+\delta_{\tau'_{a},\tau'_{b}})]^{1/2}}
\times \nonumber
\end{equation}
\begin{equation}
\label{ec8}
\times \left[\langle \tau_a \tau_b l m_l | V | \tau'_a \tau'_b l' m'_l \rangle 
+ \eta \epsilon \langle \tau_a \tau_b l m_l | V | \tau'_b \tau'_a l' m'_l
\rangle \right]. 
\end{equation}

\vspace{.2cm}

\noindent
The interaction potential depends on the total spin resulting from the coupling
of the $s_{a}=s_{b}=1$ spins of the $^{3}\Sigma$ molecules, $S=0,1,2$, and 
can be represented as\cite{Tiesinga:93}:

\begin{equation}
 \label{ec9}
V(\vec{R},\hat{r}_{a},\hat{r}_{b})=\sum_{S=0}^{2}\sum_{M_{S}=-S}^{S}V_{S}(\vec{R},\hat{r}_{a},
\hat{r}_{b})  |SM_{s}\rangle\langle SM_{s}| 
\end{equation}

\vspace{.2cm}

\noindent
where $M_{S}$ is the projection of the total spin, $M_S=m_{s_a}+m_{s_b}$. We use this
representation in order to include directly the singlet, triplet and quintet
{\em ab  initio} PESs of Ref.\cite{abi-PES} (an alternative approach was
followed in Ref.\cite{paper-Krems} since the Perugia PES is given as a sum of
a spin-independent and a spin-dependent contribution\cite{Perugia-PES}). 
In this way, matrix elements of Eq.\ref{ec8} can be further developed as

\begin{equation}
\langle \tau_a \tau_b l m_l | V | \tau'_a \tau'_b l' m'_l \rangle = 
\delta_{M_S,M'_S} \sum_{S=0}^2 (2 S + 1)
\nonumber 
\end{equation}
\vspace{-.25cm}
\begin{equation}
\times
\small{\threejm{1}{m_{s_{a}}}{1}{m_{s_{b}}}{S}{-M_S}
\threejm{1}{m_{s'_{a}}}{1}{m_{s'_{b}}}{S}{-M'_S}}
\nonumber 
\end{equation}
\vspace{-.25cm}
\begin{equation}
\label{ec10}
\times  \langle
n_{a}m_{n_{a}}n_{b}m_{n_{b}}lm_{l}|V_{S}|n'_{a}m_{n'_{a}}n'_{b}m_{n'_{b}}l'm_{l'}\rangle,  
\end{equation}

\vspace{.2cm}

\noindent 
where $(:::)$ are 3-$j$
symbols. An explicit expression for $\langle
n_{a}m_{n_{a}}n_{b}m_{n_{b}}lm_{l}|V_{S}|n'_{a}m_{n'_{a}}n'_{b}m_{n'_{b}}l'm_{l'}\rangle$  
is given in Eq.18 of Ref.\cite{paper-Krems}.

Close-coupled equations (Eq.\ref{ec7}) are solved by means a log-derivative
method\cite{Mano,Hybridprop} and using the basis set of Eq.\ref{ec4} in which, as
mentioned above, the asymptotic Hamiltonian is not diagonal.
At the point of imposing scattering boundary conditions and
thus, obtaining the scattering $S$-matrix, it is necessary to transform  
 to a new basis set $\psi^{\eta}_{\zeta_{a}\zeta_{b},l,m_l}$ giving the
 eigenstates of the fragments. For each $l,m_l$ block:

\begin{equation}
\left[ \hat{H}_{a} + \hat{H}_{b} \right]
\psi^{M\eta\epsilon}_{\zeta_{a}\zeta_{b} l m_l} = (\varepsilon_{\zeta_a} +
\varepsilon_{\zeta_b}) \psi^{M \eta \epsilon}_{\zeta_{a}\zeta_{b} l m_l},
\label{ec11}
\end{equation}

\vspace{.2cm}

\noindent
where $\varepsilon_{\zeta_{\alpha}}$ is the Zeeman fine structure energy level of
molecule $\alpha$. An unitary transformation of the log-derivative matrix onto
the new basis is performed at the end of the propagation, and then scattering
$S$-matrices and transition $T$-matrices are obtained in a standard
way\cite{paper-Krems}. The integral cross section for a
transition $\zeta_{a}\zeta_{b}\rightarrow \zeta'_{a} \zeta'_{b}$ within a
given $(M,\eta,\epsilon)$ block is finally given as 

\begin{eqnarray}
\label{ec12}
\sigma^{M \eta \epsilon}_{\zeta_{a}\zeta_{b}\rightarrow \zeta'_{a} \zeta'_{b}}
=\frac{\pi
  \left(1+\delta_{\zeta_{a},\zeta_{b}}\right)}{k_{\zeta_{a}\zeta_{b}}^{2}} \hspace{-.2cm}
 \sum_{lm_{l} l'm_{l'}} \hspace{-.2cm}
|T^{M \eta \epsilon}_{\zeta_{a}\zeta_{b}lm_{l};\zeta'_{a}\zeta'_{b}l'm_{l'}}|^{2},
\end{eqnarray}

\vspace{.2cm}

\noindent
where $T$ is the transition matrix and $k_{\zeta_{a}\zeta_{b}}^{2}/(2\mu)=
 E- \varepsilon_{\zeta_a} + \varepsilon_{\zeta_b}$ is the translational
energy of the initial channel. In obtaining Eq.\ref{ec12}, integration of the
differential cross section has been restricted over half-space for final
states satisfying $\zeta'_{a} = \zeta'_{b}$ (see Ref.\cite{paper-Krems}). 
This is  equivalent to dividing by two the cross sections
integrated over full-space to avoid double counting when the state of the
outgoing molecules is the same\cite{ourjpca, Curtiss58}.

\begin{figure}[t]
\includegraphics[width=8.8cm,angle=0.]{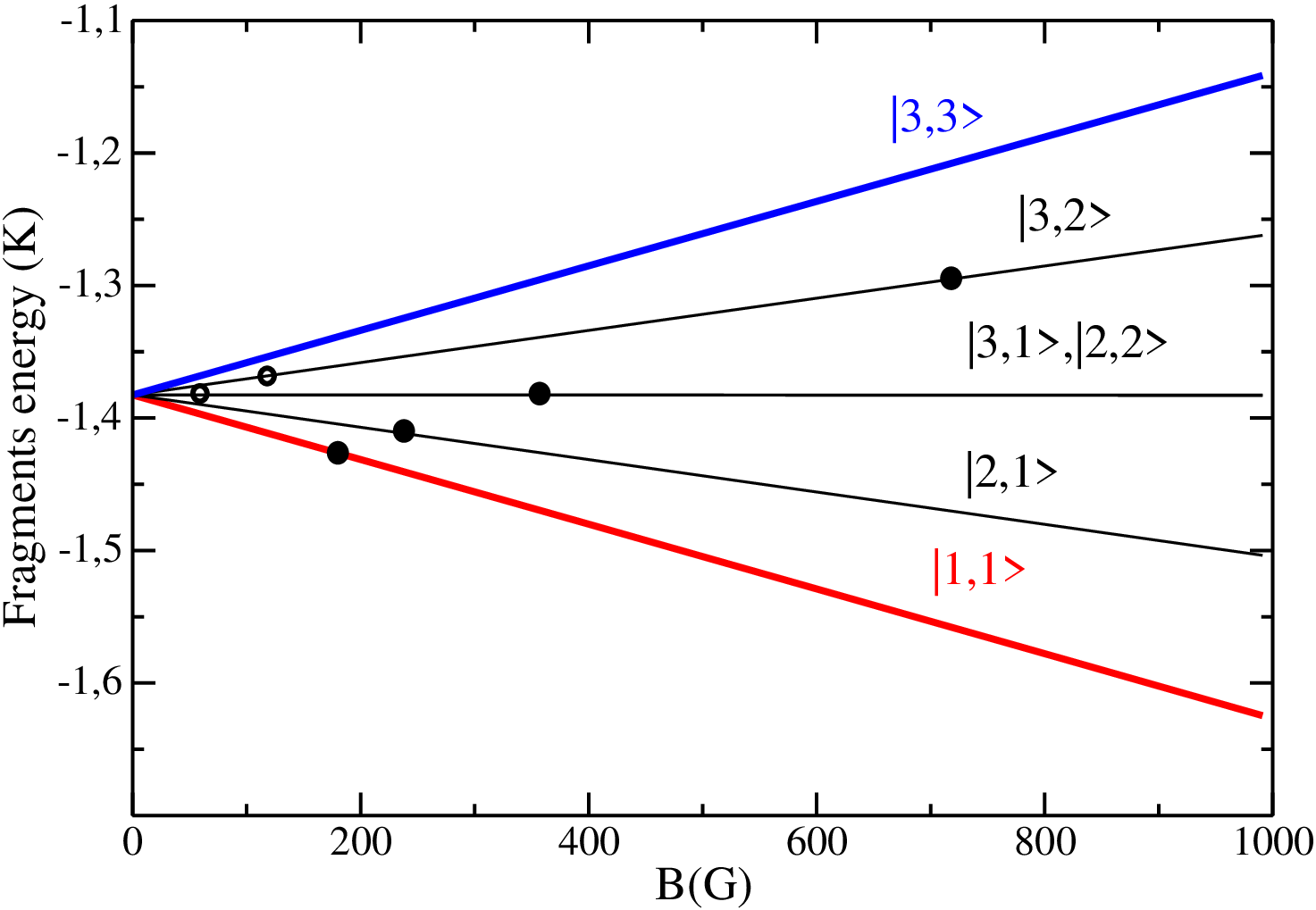}
\vspace*{.2cm} 
\caption[]{Internal energies of $^{17}O_2(\zeta_a) + ^{17}O_2(\zeta_b)$ as
functions of magnetic field. In this work, molecules are considered to be
initially in their $lfs$ states $|\zeta_a,\zeta_b\rangle=|3,3\rangle$. Open
and closed symbols indicate critical values of the field
for which $d$ and $g$ barriers, respectively, become open for the different
outgoing channels. Note also that calculations of Fig.\ref{fighfs} refer to the
{\em hfs} state $|1,1\rangle$.}    
\label{fig1}
\end{figure}

\begin{figure}[t]
\includegraphics[width=8.8cm,angle=0.]{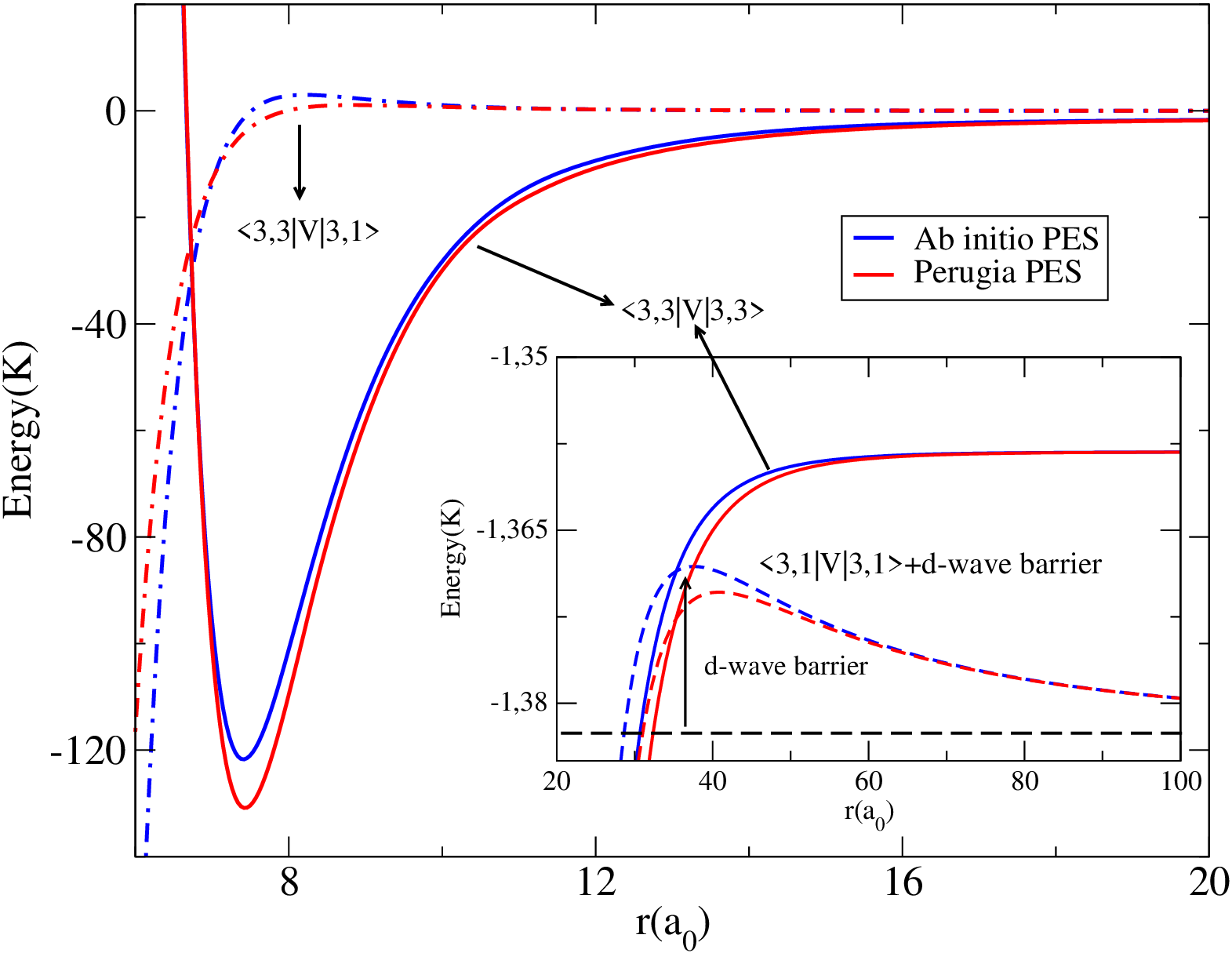}
\vspace*{.2cm} 
\caption[]{Comparison of the {\em ab initio} and Perugia potential matrix
  elements among the fragment states $|3,3\rangle$ and $|3,1\rangle$ for a
  magnetic field $B$=100 G. The long range behavior is compared in the
  inset. Note that orbital angular momentum for the entrance ($|3,3\rangle$)
  and outgoing($|3,1\rangle$)channels are 0 and 2, respectively.}    
\label{fig2}
\end{figure}

\begin{section}{Computation details}

The asymptotic Hamiltonian of Eq.\ref{ec2} is parameterized for $^{17}O_2$ by
means of accurate spectroscopic constants\cite{ref51-Krems}: $B_{e}$=1.353
cm$^{-1}$, $\gamma_{sr}$=-0.00396 cm$^{-1}$  and $\lambda_{ss}$=1.985
cm$^{-1}$. The three lowest states of the $n$= {\em even} manifold are given in
Table \ref{tableI} for a typical value of the magnetic field.
Dependence with magnetic
field of the combined $|\zeta_a,\zeta_b\rangle$ asymptotic states is depicted in
Fig.\ref{fig1}. In this work, we focus on the initial state
$|\zeta_a,\zeta_b\rangle=|3,3\rangle$, i.e., both molecules are, prior to
interaction, in their lowest {\em lfs} state. Elastic and inelastic integral
cross sections are obtained for  translational energies ranging from
10$^{-8}$ to 0.05 K.
 As we are dealing with collisions between identical (composite) bosons,
 calculations are restricted to the $\eta=+1$ block (the role of nuclear
spin can be ignored, as explained in detail in Ref.\cite{ref25-Krems}). 
Note also that to study processes involving identical internal states
(Eq.\ref{ec4}), calculations are constrained to the $\epsilon=+1$ parity (only
even $l$'s in the wavefunction expansion).

The intermolecular interaction is given by the global {\em ab initio}
PES of Bartolomei {\em et al}\cite{abi-PES}, specifically, the one referred
in that work as CC-PT2 PES.
Singlet, triplet and quintet ($S$=0,1,2)
 potentials are given\cite{abi-PES}  by the spherical harmonic expansion

\begin{equation}
 \label{ec13}
V_{S}(\vec{R},\hat{r}_a,\hat{r}_b)=(4\pi)^{3/2} \hspace{-.15cm} 
\sum_{\lambda_{a} \lambda_{b} \lambda}  \hspace{-.15cm}
V_{S}^{\lambda_{a}\lambda_{b}\lambda}(R)
A_{\lambda_{a} \lambda_{b} \lambda}(\hat{R},\hat{r}_a,\hat{r}_b),
\end{equation}

\vspace{.2cm}

\noindent

\noindent
where $A_{\lambda_{a} \lambda_{b} \lambda}$ is given as a combination of
spherical harmonics and $\lambda_{a}$, $\lambda_{b}$ and $\lambda$ are even integers
(due to the symmetry of the four identical nuclei). The radial coefficients 
$V_{S}^{\lambda_{a}\lambda_{b}\lambda}(R)$ were obtained by means of
quadratures of the supermolecular {\em ab initio} energies over the angular
variables, obtaining a total of 29 coefficients for the quintet PES and 27 for
the singlet and triplet ones. 
The PESs are extended asymptotically ($R>$ 19 bohr) using analytical functions
 (common to the three multiplicities) based on high level {\em ab
  initio} calculations of electrostatic, dispersion and induction long range
coefficients\cite{long-range}. In the following Section, we present a comparison with
calculations using the Perugia PES\cite{Perugia-PES}, which comprises
 just four radial terms (for each multiplicity) derived from a
multi-property fitting analysis. 
To give a flavor of the similarities/differences
between the two PES considered, we present in Fig.\ref{fig2} the dependence
with the intermolecular distance of the potential matrix elements among the
{\em lfs} state $|3,3\rangle$ and the (one spin flipping) relaxation channel
$|3,1\rangle$. These matrix elements are relevant to the mechanisms proposed
by Krems and Dalgarno\cite{ref20-Krems} and by Volpi and
Bohn\cite{ref24-Krems}. Note that for initial states approaching in an $s$ wave,
conservation of $M$ forbids $s$ waves in the spin relaxation channels (see
Eq.\ref{ec6} and Ref.\cite{ref24-Krems}).  It can be seen that there are some quantitative
differences in the coupling as well as in the long range behavior. A comparison of
properties related to the van der Waals (vdW) coefficient $C^{000}_6$ is
summarized in Table \ref{tableII}.

Cross sections are computed using the code developed
by Tscherbul {\em et al}\cite{paper-Krems}, modified by us to
include the hybrid log-derivative/Airy propagator of Alexander and
Manolopoulos\cite{Hybridprop}. Related routines were taken from the MOLSCAT
code\cite{Molscat}. 
In this way, the log-derivative propagator of Manolopoulos\cite{Mano} is used
in the strongly coupled region (from 4.5 $a_{0}$ to 40.8 $a_{0}$) with a
fixed short step (0.04 $a_0$), 
whereas the Airy propagator of Ref.\cite{Hybridprop} is used for the long
range region (from 40.8 $a_{0}$  to 202. $a_{0}$) with a variable step size
(the ratio between adjacent step sizes being of 1.05). 
Comparing with the
original code of Tscherbul {\em et al}, where only the log-derivative
propagator was used, we found that the errors are less than  $0.5\%$ while the new
propagation is about 10 times faster due to the smaller number of
integration steps as well as the use of the computationally less expensive
Airy propagator.

The total wave function is expanded using a basis set comprising 
three rotational levels ($n_{a}$, $n_{b}$=0, 2, 4) and four partial
waves ($l$=0, 2, 4, 6), equal to that employed in Ref.\cite{paper-Krems}. Although
exact positions of the resonances might change with an increase of the basis
size, this basis is sufficient to retrieve the main features
of the collision dynamics.
Regarding the convergence of the cross sections with the
 projection of the total angular momentum, $M$,  it is found that for translational
energies lower than  $10^{-4}$ K, just the $M=2$ block calculation is
sufficient, while for larger energies, five blocks ($M=$0-4) have to be summed
up.  For a single energy and magnetic field calculation, typical run times are
of about 18 and 90 hours, respectively.

\end{section}

\section{Results and discussion }

\vspace*{-.2cm}

We present first the results concerning the $B$-field dependence at very low
energies and, in a subsequent section, we report those related to the translational
energy dependence, including the transition from the ultracold to the cold
regimes. 

\vspace*{-.35cm}

\subsection{Magnetic field dependence at 1 $\mu$ K}
\vspace*{-.2cm} 

\begin{figure}[t]
\includegraphics[width=8.8cm,angle=0.]{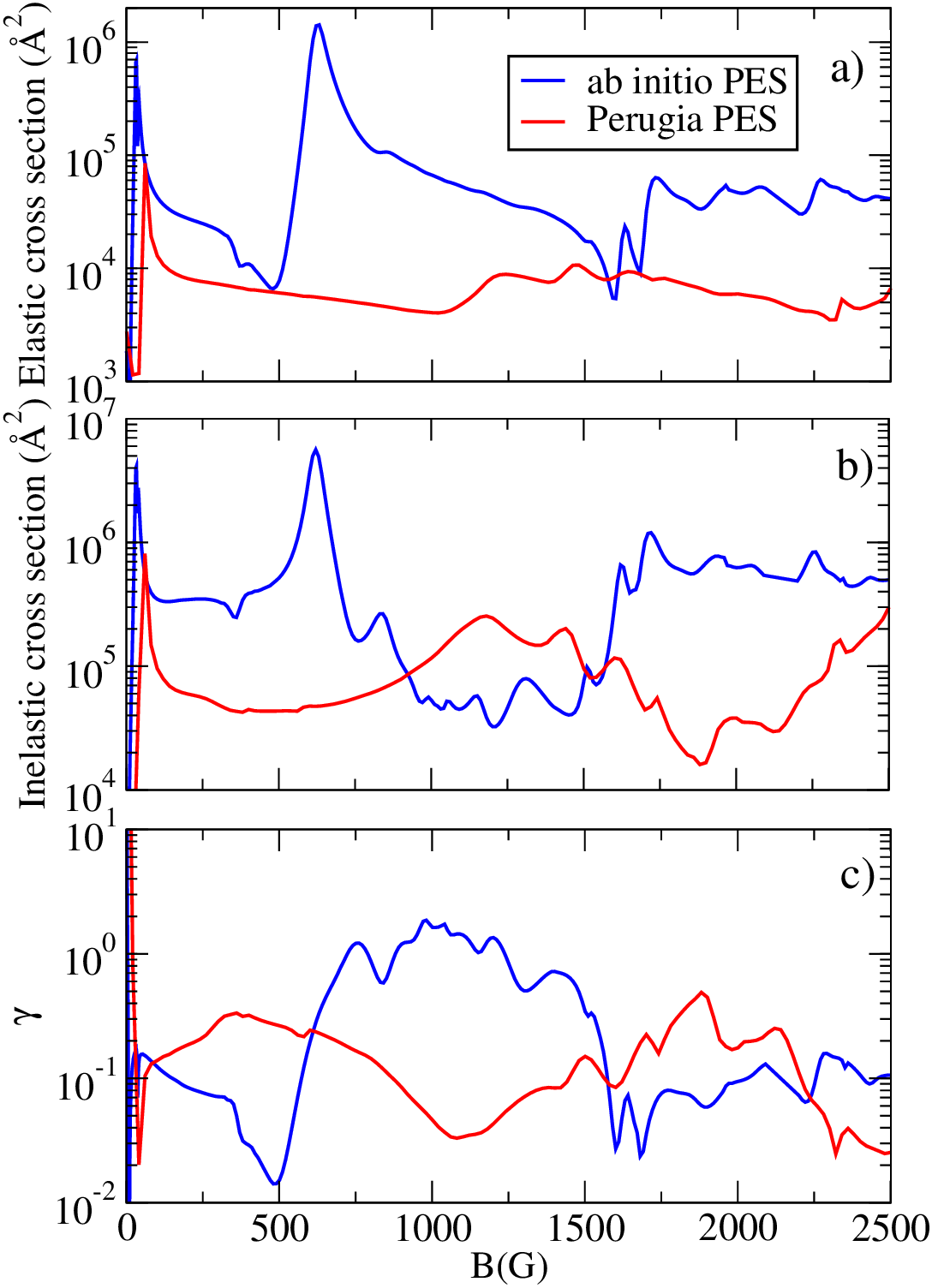}
\vspace*{0.3cm} 
\caption[]{Magnetic-field dependence for collisions of $^{17}O_2 + ^{17}O_2$
  in the initial {\em lfs}  state $|3,3\rangle$ and translational energy of 1
  $\mu$K. a) Elastic cross sections; b) Total inelastic cross sections and c)
  Ratio $\gamma$ between elastic and inelastic (untrapping) cross
  sections. Blue and red colors correspond to using the {\em ab
    initio}\cite{abi-PES} and the Perugia\cite{Perugia-PES} PESs,
  respectively.}    
\label{fig3}
\vspace*{-0.3cm}
\end{figure}

The magnetic-field dependence of the {\em ab initio} and Perugia cross sections
for the {\em lfs} state $|3,3\rangle$ at 1 $\mu$K is summarized in
Fig.\ref{fig3} (panels a and b). In Fig.\ref{fig3}.c we report the
elastic-to-inelastic ratio, $\gamma$, more specifically,
the ratio between the elastic cross section and those inelastic ones leading
to  untrapped states: $|\zeta'_a,\zeta'_b\rangle=|3,1\rangle,|2,2\rangle,
|2,1\rangle$ and $|1,1\rangle$. Note that new calculations with the Perugia
PES were performed using the same basis set as with the {\em ab initio} PES
(there are some quantitative changes between present calculations and those
given in Fig.3 of Ref.\cite{paper-Krems} were a smaller basis was employed). There are
various noticeable differences between the two PESs. 
On the one hand, {\em ab initio} elastic and inelastic cross sections
(Figs.\ref{fig3}.a,b) are much larger than the Perugia ones and,
in addition, they exhibit more  marked Feshbach resonance structures.
On the other hand, although there are large variations of the
elastic-to-inelastic ratio with the magnetic field, it can be seen that
both PESs produce values which, on average, are of the same order of
magnitude. The cases of very low fields ($B<$ 50 G), where $\gamma$ is much
larger for the Perugia PES, and around 1000 G, where the {\em ab initio} value
becomes very large, are discussed in more detail below.

We discuss first the background behavior of the cross sections of
Fig.\ref{fig3}. The elastic cross sections correspond to a background
scattering length, $a_{bg}$, of about 118 and 32 $a_0$ (in absolute value),
for the {\em ab initio} and Perugia PESs, respectively. These quantities are
larger than the scattering lenghts purely due to the vdW
potential\cite{Gribakin:93}, $\overline{a}$, of 22 and  24 $a_0$, respectively.   
The particularly large value of the {\em ab initio} elastic cross section 
can be explained by existence of a close quasibound state varying
 with magnetic field at the same rate than the entrance channel. 
Regarding inelastic cross sections, the {\em ab initio} one is on average
about  10 times larger than the Perugia result.
This difference can be qualitatively rationalized by resorting to the analytic van
der Waals theory\cite{Julienne:06,Gao:98b}, which takes the solutions of the vdW
potential\cite{Gao:98a} as the reference for the multichannel quantum
defect theory\cite{Julienne:89}. A key parameter in that approach is the
short range squared amplitude of the entrance channel wave function, which
near threshold is proportional to\cite{Julienne:06}    

\begin{equation}
\lim_{k_0\rightarrow 0} C_{bg}(k_0)^{-2} = k_0
    \overline{a} \left[ 1+\left(1-\frac{a_{bg}}{\overline{a}}\right)^2 \right],
\label{ec15}
\end{equation}

\noindent
$k_0$ being the wavenumber of the incoming channel. Since
inelastic cross sections are proportional to
$C_{bg}(k_0)^{-2}$\cite{Julienne:89},  Eq.\ref{ec15} implies that the 
value of $a_{bg}$ affects the threshold behavior of the inelastic cross
sections. It follows, then, that the very large {\em ab initio}
inelastic cross sections are explained by the magnitude of the corresponding
background
scattering length. Within this framework, one can expect that the
elastic-to-inelastic ratio becomes less sensitive to $a_{bg}$ than the cross
sections themselves, since both elastic and inelastic cross sections are
approximately proportional to $a_{bg}^2$. This is the result of
Fig.\ref{fig3}.c, where the average value of $\gamma$ is about the same for
both potentials.

\begin{figure}[t]
\includegraphics[width=8.8cm,angle=0.]{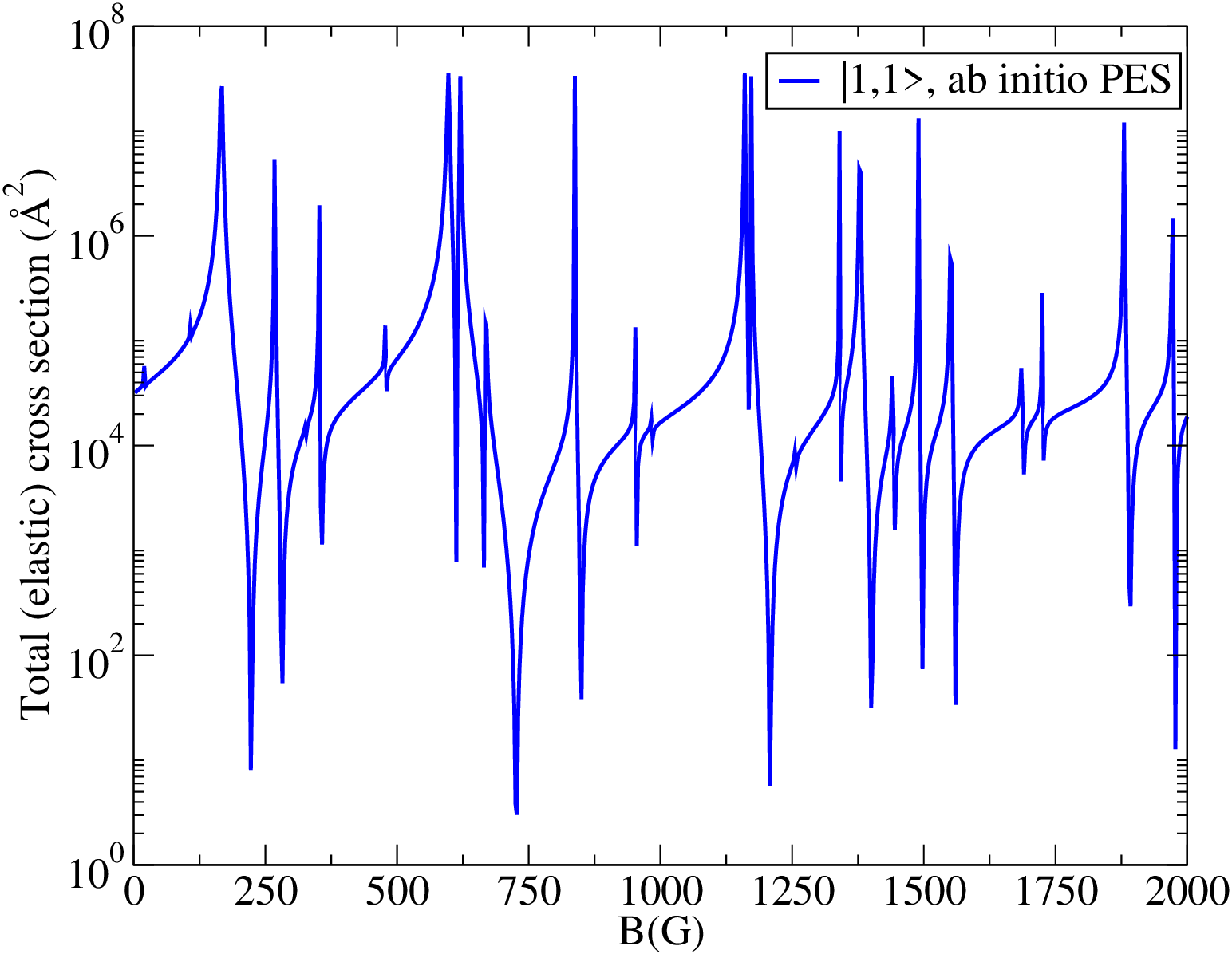}
\vspace*{-0.2cm} 
\caption[]{Total cross section vs. magnetic-field for 
  the {\em hfs}  state $|1,1\rangle$ at a translational energy of 1
  $\mu$K and using the {\em ab initio} PES. Note that only the elastic
  channel is open. The result can be compared with Fig.4 of
  Ref.\cite{paper-Krems}, corresponding to the Perugia PES.}   
\label{fighfs}
\vspace*{-.2cm} 
\end{figure}

We now turn to discuss the resonant structures 
 of Fig.\ref{fig3}. At this point, it is convenient to mention
 the work of Hutson\cite{Hutson:07} who analyzed the threshold behavior of
 Feshbach  resonances in the presence of inelastic scattering. He found
 that -in contrast to the case of a pure elastic scattering- resonance peaks
 may be significantly suppressed and, in this way,
 the collisional process may become insensitive to the details of the
 potential. With this in mind, the profiles obtained in Fig.\ref{fig3} are 
 rather unexpected given the considerable anisotropy of the $O_2-O_2$ interaction. 
In connection with this issue, let us digress for a while and study 
the resonance patterns for a purely elastic scattering event, as is the
case of  the magnetic field dependence of the lowest {\em high field seeking}
({\em hfs}) state $|1,1\rangle$ (see Fig.\ref{fig1}). The result for the {\em
  ab initio} PES at 1  $\mu$K, using a a reduced basis ($n_{max}$=4,$l_{max}$=4), is
shown in Fig.\ref{fighfs} and can  be directly compared with Fig.4 of
Ref.\cite{paper-Krems}.  For both PESs, a high density of very pronounced
resonances is obtained. For the {\em ab initio} PES there is a slightly larger
number of peaks, and some of them are wider. Also, the baseline
of the {\em ab initio} cross section is
much larger than the Perugia one, as occurs for the {\em lfs} state. 
 A similar density of quasibound states is expected when the entrance channel 
is the $lfs$ state, but presence of inelastic channels substantially modify 
 the resonance lineshapes\cite{Hutson:07}. To show this, it is convenient to
write down the behavior of the $S$ matrix in the neighborhood of an isolated
resonance\cite{Feshbach:58,Hutson:07},

\begin{equation}
S_{jk}(E) = S^{bg}_{jk} - i \frac{g_{Ej}g_{Ek}}{E-E_r + i \Gamma_{E}/2},
\label{ec16}
\end{equation}

\noindent
where $k$ and $j$ are the incoming and outgoing channels, respectively,
$S^{bg}_{jk}$ is the background $S$ matrix, $E$ is the total energy, $E_r$ is
the resonance position, $\Gamma_{E}$ is the resonance
width and (complex) $g_{Ei}$ involve couplings between resonance and 
channel $i$ wavefunctions\cite{Miller:70}, such that the partial width for
channel $i$ is given as $\Gamma_{Ei}=|g_{Ei}|^2$ and  $\Gamma_{E}= \sum_i
\Gamma_{Ei}$. A key point in Hutson's argument is that $g_{Ek}$ elements are
proportional to the square root of the incoming channel wavenumber
$k_0^{1/2}$. Then, as $k_0$ decreases and if the resonant state is also
coupled to inelastic channels, the radius of the circle
described by $S_{jk}$ drops to zero and peaks in cross sections become
significantly supressed\cite{Hutson:07}. The analytical vdW theory gives a more
detailed threshold behavior of the $g_{Ek}$ elements, as they become proportional to
the square root of Eq.\ref{ec15}. Hence, if $a_{bg}$ is sufficiently large,  
$g_{Ek}$ will tend its threshold value (zero) rather slowly, and as a
consequence, more pronounced peaks in the cross sections can be obtained. This
explains why we find a marked resonance structure, especially for the {\em ab
  initio} PES. Nevertheless, as noted in Ref.\cite{Hutson:07}, a relatively
large ratio between elastic and inelastic partial widths is also needed in
order to obtain pronounced resonance profiles. It is reasonable to expect that,
 among all the
    quasibound states that should be crossing the $lfs$ state, only some of them
    will have particularly large {\em elastic} partial widths, so only a few
    marked resonance features will ``survive'', as in fact it occurs (Fig.\ref{fig3}).

\begin{figure}[t]
\includegraphics[width=8.8cm,angle=0.]{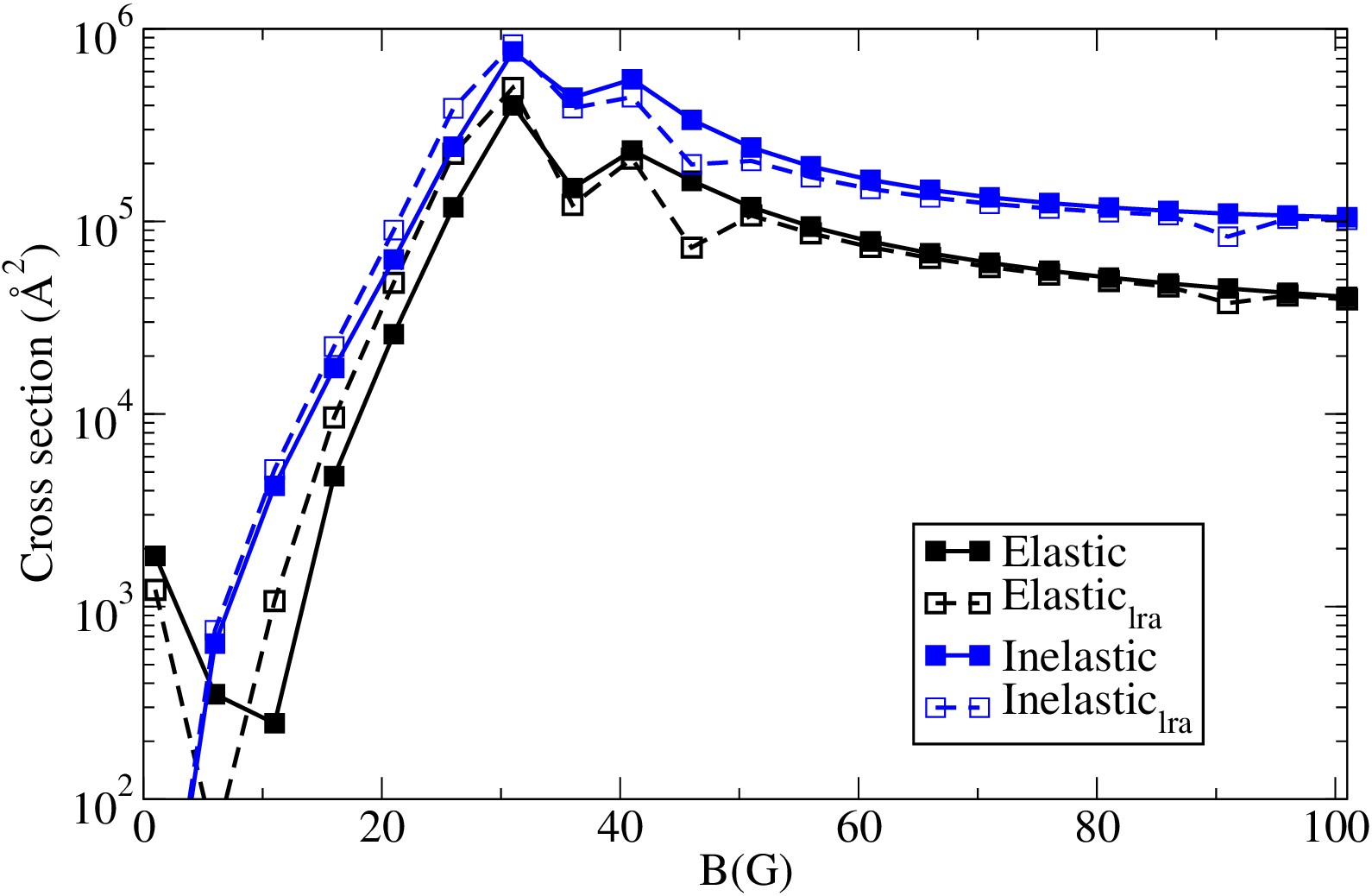}
\vspace*{-0.2cm} 
\caption[]{Effect of the long-range anisotropy of the {\em ab
    initio} PES: Magnetic-field dependence of elastic and total inelastic
  cross sections for the {\em lfs}  state $|3,3\rangle$ at 10
  $\mu$K. Thick lines joined by filled squares show results using the correct long
  range anisotropy\cite{abi-PES,long-range}, while dashed lines joined by open
  squares correspond to calculations where the long range anisotropy of the
  interaction has been switched off.}   
\label{fig5}
\vspace*{-.2cm} 
\end{figure}

We have just seen that a large $a_{bg}$ enhances the short range couplings
between the resonance and the incoming wavefunctions. In this situation the
dynamics must become very sensitive to the short range region of the potential. 
 In order to study the role
played by the short range vs. the long-range features of the intermolecular
potential, we have performed a test calculation where the long-range
anisotropy of the potential is switched off. To this end, the {\em ab initio} PES has been
modified by imposing, for $R>$ 19 $a_0$, an exponential decay of all 
radial terms of Eq.\ref{ec13} except the isotropic one
$(\lambda_a,\lambda_b,\lambda)= (0 0 0)$. The new cross sections are compared
with those corresponding to the correct long range behavior in
Fig.\ref{fig5}. This figure clearly shows that the resonance structure is
rather insensitive to the long-range anisotropy of the
interaction and, therefore, short range couplings must be playing a dominant
role.

Finally, it is interesting to note from Fig.\ref{fig3}.b that, for the  {\em
  ab initio} PES, there is a significant suppression of inelastic
scattering for  magnetic fields ranging from 750 to 1500 G. This feature must
be related with the prominent resonance at about 600 G and it must be due to
interferences between the background and resonant $S$ matrices leading to
asymmetric line-shapes of the state-to-state cross 
sections\cite{Fano:61}. Note that this reduction entails
 a considerable increase of the ratio $\gamma$ 
for a wide range of magnetic fields. A similar behavior (with an even 
larger suppression of inelastic scattering) has been
found in $^4$He + $^{16}$O$_2$ magnetic Feshbach resonances\cite{Hutson:09}. 

Analogously, it is also worth mentioning that, for the {\em ab initio} results,
the elastic scattering on the left-hand-side of the resonance at about $B$= 30
G is  suppressed. This feature, already present in Fig.\ref{fig3}, can be more
clearly seen in Fig.\ref{fig5}, where
the {\em ab initio} elastic cross section becomes very small around 10 G. In
this case, the corresponding ratio $\gamma$ becomes much smaller than 
expected (from the well known effect of suppresion inelastic scattering due to
centrifugal barriers\cite{ref25-Krems,ref24-Krems,paper-Krems}).

\subsection{Translational energy dependence}

\begin{figure}[t]
\includegraphics[width=8.8cm,angle=0.]{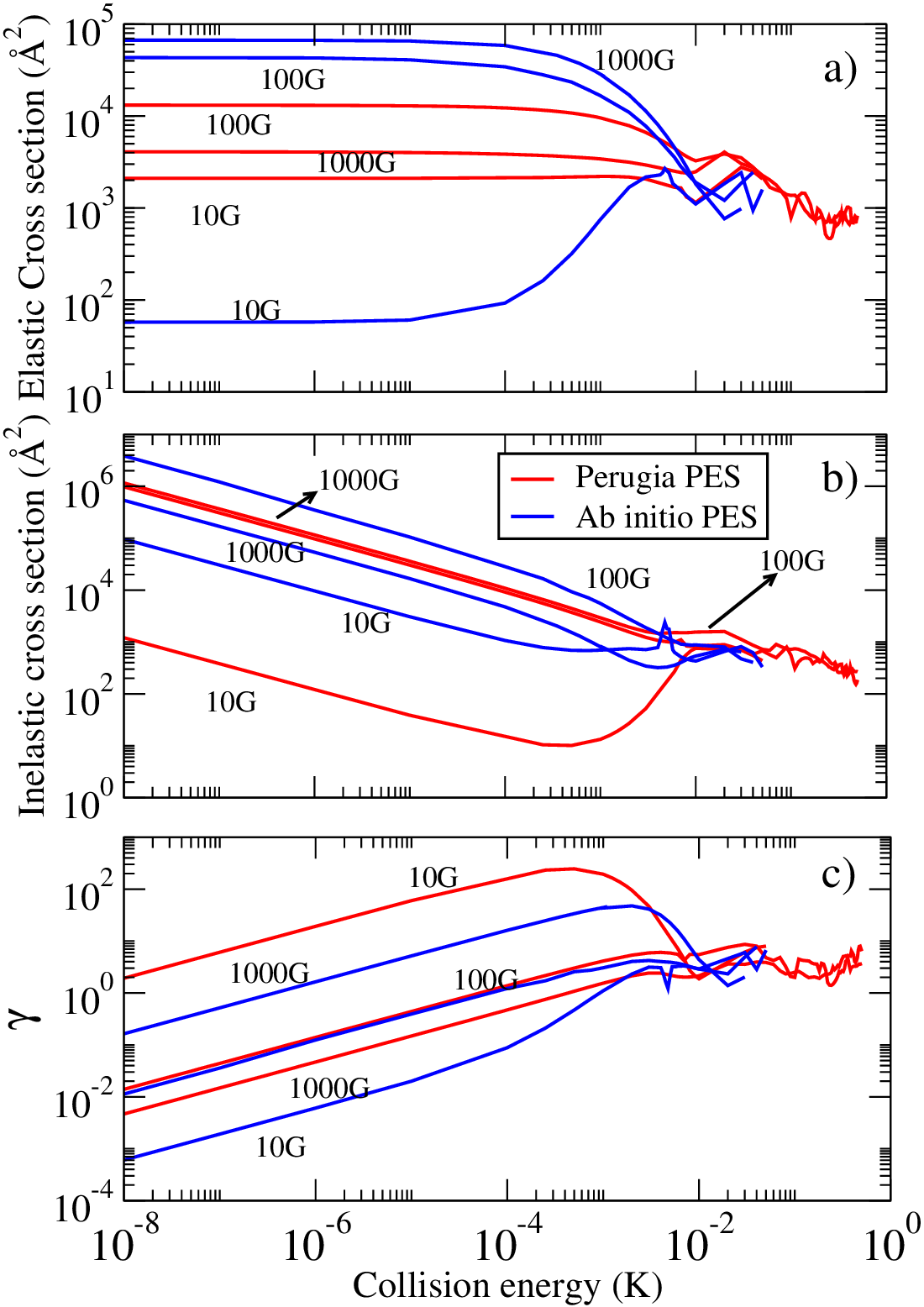}
\vspace*{-0.2cm} 
\caption[]{Translational energy dependence of the collisional processes for
  the {\em lfs}  state $|3,3\rangle$: a comparison between {\em ab initio} and
  Perugia PESs for different values of magnetic field. a) Elastic cross sections; b) Total
  inelastic cross sections and c)   Elastic-to-inelastic ratio $\gamma$.}   
\label{figdepE}
\vspace*{-.2cm} 
\end{figure}

In Fig.\ref{figdepE}, dependence of the cross sections with
kinetic energy is given for several selected values of the magnetic field. In
agreement  with predictions based on the analytical
vdW theory\cite{Julienne:06}, two very different regimes are
noticed for energies larger or smaller than $E_{vdW} \approx$ 10 mK (see 
Table \ref{tableII}).
For the higher energy range, elastic and inelastic cross sections exhibit a
weak dependence with the field, the Perugia ones
being larger than their {\em ab initio} counterparts, in consistency with
previous studies at higher energies\cite{ourjpca}. For
energies lower than the crossover ($E_{vdW}$), cross sections become more
dependent on the magnetic field. This is mainly due to the effect of the
resonances in the ultracold regime, but in the case of the Perugia PES,
supression of inelastic cross
sections at low fields (due to the centrifugal
barriers\cite{paper-Krems,ref25-Krems}) also plays a role. 

It is interesting to highlight that a relatively high value of the
elastic-to-inelastic ratio has been obtained between 1 and 10 mK in the {\em ab
  initio} calculation at 1000 G (Fig.\ref{figdepE}.c). This result is related
to the asymmetry of the lineshape and the suppression of spin-changing 
processes on the right-hand-side of the resonance at 600 G and 1 $\mu$ K,
discussed above (Fig.\ref{fig3}).

\begin{figure}[t]
\includegraphics[width=8.8cm,angle=0.]{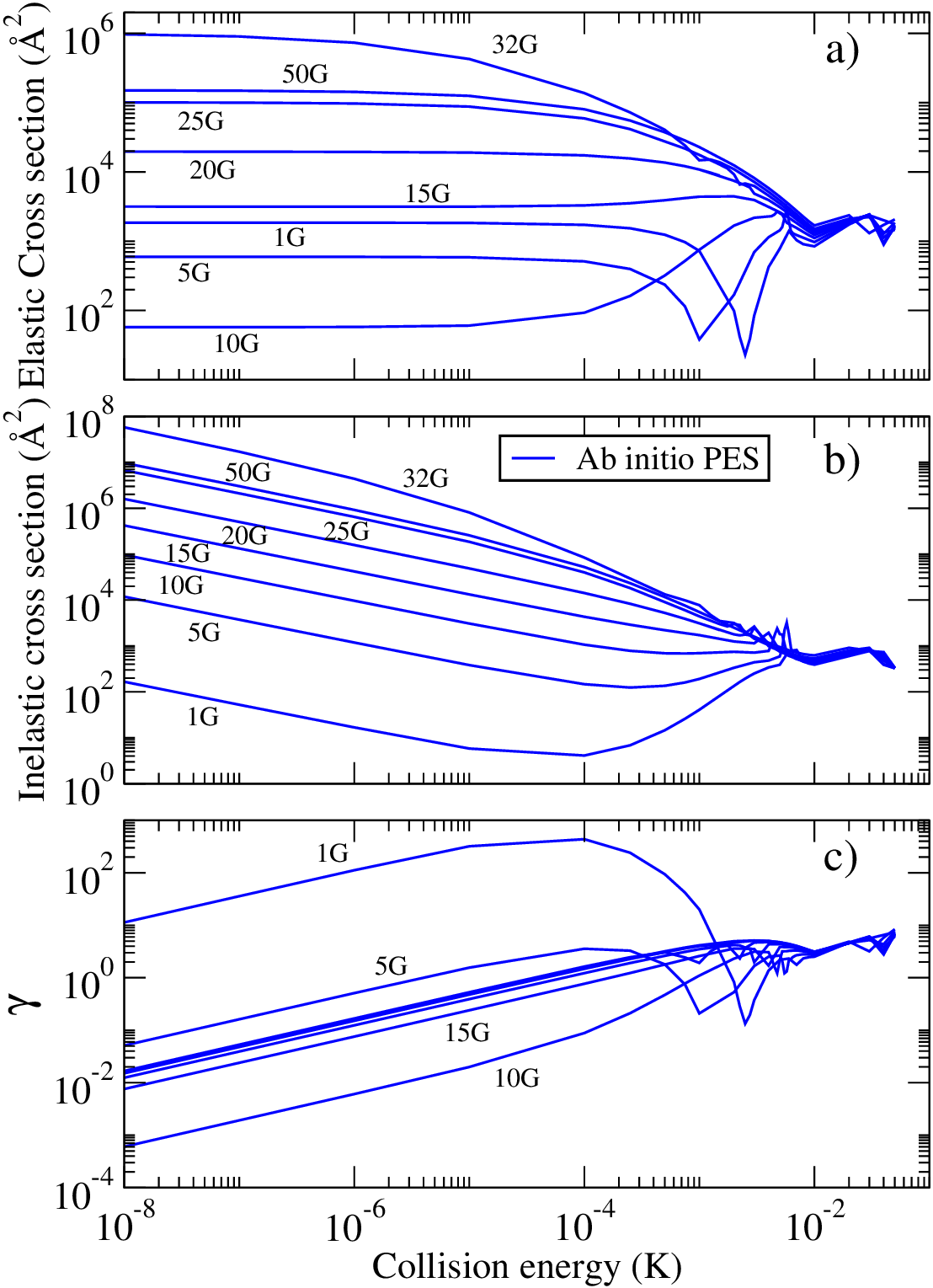}
\vspace*{-0.2cm} 
\caption[]{Same as Fig.\ref{figdepE} but for  different values of magnetic field
   near the 32 G resonance of the calculations with the {\em ab
    initio} PES.}   
\label{figres}
\vspace*{-.2cm} 
\end{figure}

A more detailed study of the {\em ab initio} cross sections for low
values of the field ($B\le$ 50 G) 
is given in Fig.\ref{figres}. An impressive dependence with
$B$ is noticed for energies just below 10 mK. Between 1 and 10 mK,
complicated resonance structures are seen which are particularly acute for 
the elastic cross section. These features are related to the
prominent resonance around 30 G at much lower energies (reported in 
Fig.\ref{fig3} and more clearly seen in Fig.\ref{fig5}). In other words, they are 
expressions -at several different energies and magnetic fields- of the same
quasibound state. For instance, note the
resemblance between the asymmetric line shapes of the elastic cross section
at $B$= 1 and 5 G and between 1 and 10 mK (Fig.\ref{figres}.a), with the magnetic
field dependence at much lower energies for fields $B<$ 30 G, shown in Fig.\ref{fig5}. A 
detailed tracking of these resonances would involve non-trivial lineshape fittings 
 and has not been attempted here. On the other hand, it should
be noted that, for the range of magnetic fields of Fig.\ref{figres} and up to
translational energies of at least 1 mK, spin-changing collisions
should be suppressed due to existence of centrifugal barriers in all 
outgoing channels\cite{paper-Krems,ref25-Krems}. In Fig.\ref{figres}.b it can
be seen that, except for the lowest value of $B$ (1 G), such a suppression does not
occur, in contrast with the results using the Perugia
PES (see Ref.\cite{paper-Krems} and Fig.\ref{figdepE}.b). This must be due to
a significant tunneling through the centrifugal barriers for energies/fields
close to the resonance. Consequently, the ratios $\gamma$ are particularly
small for this range of fields (Fig.\ref{figres}.b).

\begin{figure}[t]
\includegraphics[width=8.8cm,angle=0.]{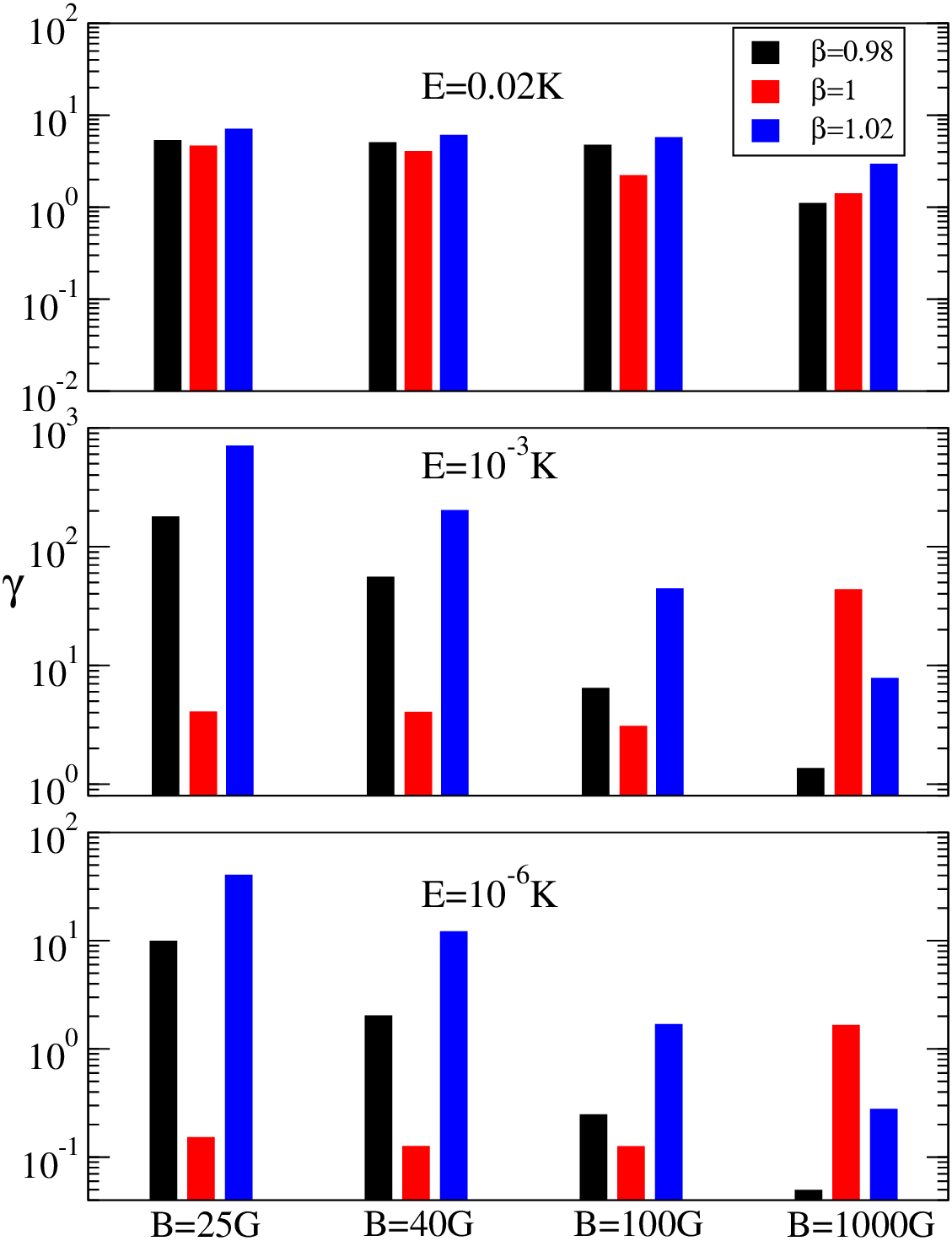}
\vspace*{0.3cm} 
\caption[]{Study of the sensitivity of the elastic-to-inelastic ratio $\gamma$ to
  the anisotropy of the {\em ab initio} PES. All terms of the spherical
  harmonic expansion are multiplied by $\beta$ except the isotropic one. The
  effect is shown for different translational energies and magnetic fields.}   
\label{fig8}
\vspace*{-.5cm} 
\end{figure}

A further analysis of the sensitivity of the elastic-to-inelastic ratio to the
details of the PES has been performed. We have artificially modified the
anisotropy of the present {\em ab initio} PES by multiplying all the terms in
the spherical harmonic expansion (Eq.\ref{ec13}) -except the isotropic ones-
 by a factor $\beta$ ranging form 0.98 to
1.02. In Fig.\ref{fig8} we show the results for different translational
energies and magnetic fields. In can be seen that, while for 20 mK there is
not a strong variation of $\gamma$ with $\beta$, for lower
energies (1 mK and 1 $\mu$K), this ratio changes tremendously with the
anisotropy of the potential. In the new calculations ($\beta=$ 0.98 and 1.02),
no nearby resonances appear for the energies/fields considered and hence,
results are ``more standard'', i.e., very large values of $\gamma$
are now attained for low values of the field ($B<$ 50 G), in agreement with
the expected suppression of inelastic scattering, but smaller $\gamma$'s are
obtained for $B$= 1000 G.  However, note that, contrarily to a first order
perturbation theory, the largest ratios are obtained with the most
anisotropic PES ($\beta$= 1.02).

\section{Concluding discussion}

We have performed a detailed study of cold and ultracold molecule-molecule
collisions in the presence of a magnetic field for a system with a
significant anisotropy such as $O_2$+$O_2$.   A thorough comparison has been
made between a high  quality {\em ab initio} PES and previous
studies\cite{paper-Krems} where a a different PES was used.
Several interesting findings have emerged from this approach regarding the
anisotropy as well as the relative influence of long and short components of
the interaction. For the {\em ab initio} PES, a large background scattering
length gives rise to pronounced resonance structures in the 
ultracold regime (translational energies $<$ 10 mK). As a consequence, the
ratio between elastic
and inelastic cross sections, $\gamma$, is very dependent on the
magnetic field as well as on the short range anisotropy of the PES.
 Therefore,
quantitative predictions for this important parameter become rather risky.
However and as a general trend, we can indicate that high values of $\gamma$ could be
 achieved in the vicinity of asymmetric Fano resonances, or for low fields, $B$
$<$ 50 G. Note that the maximum temperature that can be held in a trap with
such a depth would be of about 1 mK\cite{paper-Krems,Doyle:95}. 

A key issue is the large density of quasibound states of the $O_2$+$O_2$
system, best illustrated in the magnetic field dependence of the elastic cross
sections of the lowest high field seeking state. In view of this,
having obtained a large background scattering length does not seem a rare event.
Present behavior might be characteristic of 
a range of molecule-molecule systems as well, that is to say, as the number of degrees of freedom
increases, a larger density of quasibound states, including near threshold
resonances, can be expected\cite{Bohn:02}, which in turn makes dynamics
richer.  Very recently, Suleimanov and 
Krems\cite{Suleimanov:11} have proposed an efficient method for locating
Feshbach resonances in external fields. The new
method could be very useful for the comparison of spectral patterns 
obtained from different potentials or between different molecular
systems.

\section{Acknowledgments}

We are indebted to Roman V. Krems for encouragement and for giving us essential
insight along several stages of this work. We wish to thank
M. H. Alexander, D. E. Manolopoulos and J. M. Hutson for the use of the Hybrid
Propagator routines of the MOLSCAT code, and to M. Bartolomei,
E. Carmona-Novillo and R. Hern{\'a}ndez-Lamoneda for the use of the {\em ab
  initio} PES. J.P.-R. acknowledges hospitality in the Deparment of Chemistry
of UBC (Canada) and support from a predoctoral JAE CSIC grant.
The work has been funded by Ministerio de Ciencia e Innovaci{\'o}n  
 (Spain, grants CTQ2007-62898-BQU and FIS2010-22064-C02-02).  
We also thank CESGA (Spain) for allocation of computing time.

\vspace{.3cm}

 
\bibliography{ultracold.bib}

\newpage

\begin{table}
\caption{Energies (in K) and coefficients (in the basis of
  Eq.\ref{ec5}) of the three lowest states of  $^{17}O_{2}$ (Eq.\ref{ec2}) for a magnetic field
  $B$= 100 G.}  
\centering                            
\begin{tabular}{rrrrrrrrr}              
\hline
 & & $\varepsilon_{\zeta}$ (K) &  & $C_{\tau \zeta}$  & & $|n$ & $m_n$ & $m_s\rangle$ \\
\hline                                      
 $\zeta=1$ & & -0.7035 & &  0.9687 & & $|0$ &  0 & $-1\rangle$\\
            &     &    & & -0.1922 & & $|2$ & -2 & $1\rangle$\\
            &     &    & &  0.1361 & & $|2$ & -1 & $ 0\rangle$\\
            &     &    & & -0.0786 & & $|2$ &  0 & $-1\rangle$\\
\hline
 $\zeta=2$ & & -0.6913 & &  0.9686 & & $|0$ & 0 & $0\rangle$\\
      &  &             & & -0.1361 & & $|2$ & -1 & $1\rangle$\\ 
           &  &        & &  0.1573 & & $|2$ & 0 & $0\rangle$\\
      &     &          & & -0.1364 & & $|2$ & 1 & $-1\rangle$\\
\hline
 $\zeta=3$ & & -0.6791 & & 0.9684 & & $|0$ & 0 & $1\rangle$\\
       &     &         & & -0.0788 & & $|2$ & 0 & $1\rangle$ \\ 
       &     &         & & 0.1364 & & $|2$ & 1 & $0\rangle$\\
       &    &          & & -0.1931 & & $|2$ & 2 & $-1\rangle$ \\
\hline
\end{tabular}
\label{tableI}
\end{table}

\newpage

\begin{table}
\caption{Parameters associated to the long range behavior of the
  {\em ab initio} and Perugia potentials: isotropic vdW 
  coefficient $C_6^{000}$, scale length ($R_{vdW}$) and energy ($E_{vdW}$) of
  the analytical vdW theory\cite{Julienne:06}, and height of the $d$-wave centrifugal
  barrier. $B_{min}$ is the critical magnetic field for which the  $|3,3\rangle
  - |3,1\rangle$ Zeeman splitting becomes larger than the $d$-wave barrier.}  
\centering                            
\begin{tabular}{lllll}              
\hline
 & & {\em ab initio}  &  & Perugia \\ 
\hline                                      
$C_6^{000}$(a.u.)     & & 62.39 & & 88.70 \\
$R_{vdW}(a_0)$       & & 22.17 & & 24.21 \\
$E_{vdW}$ (mK)  & & 10.4  & & \hspace*{0.1cm} 8.7 \\
$E_0(l=2)$ (mK) & & 14.7 & & 12.3 \\
$B_{min}$ (G)   & & 55    & & 46 \\
\hline
\end{tabular}
\label{tableII}
\end{table}

\newpage

\newpage

\newpage

\newpage

\newpage

\newpage

\newpage

\newpage

\end{document}